\documentclass[twocolumn]{aastex631}

\usepackage{graphicx}
\usepackage{amssymb}
\usepackage{amsmath}

\makeatletter
\DeclareRobustCommand{\HI}{%
  \mbox{H\check@mathfonts\fontsize\sf@size\z@\selectfont I}%
}
\makeatother

\makeatletter
\DeclareRobustCommand{\HII}{%
  \mbox{H\check@mathfonts\fontsize\sf@size\z@\selectfont II}%
}
\makeatother

\accepted{February 12, 2023}

\shorttitle{The effect of stellar mass and halo mass on the assembly histories of satellite galaxies}
\shortauthors{Oyarz\'un et al.}

\graphicspath{{./}{figures/}}

\begin{document}

\title{\textbf{SDSS-IV MaNGA: \\ The effect of stellar mass and halo mass on the assembly histories of satellite galaxies}}

\email{goyarzun@ucsc.edu}

\author{Grecco A. Oyarz\'un}
\affiliation{Astronomy Department, University of California, Santa Cruz, CA 95064, USA}
\affiliation{The William H. Miller III Department of Physics \& Astronomy, Johns Hopkins University, Baltimore, MD 21218, USA}

\author{Kevin Bundy}
\affiliation{University of California Observatories - Lick Observatory, University of California, Santa Cruz, CA 95064, USA}

\author{Kyle B. Westfall}
\affiliation{University of California Observatories - Lick Observatory, University of California, Santa Cruz, CA 95064, USA}

\author{Ivan Lacerna}
\affiliation{Instituto de Astronom\'ia y Ciencias Planetarias, Universidad de Atacama, Copayapu 485, Copiap\'o, Chile}
\affiliation{Millennium Institute of Astrophysics, Nuncio Monsenor Sotero Sanz 100, Of. 104, Providencia, Santiago, Chile}
	
\author{Renbin Yan}
\affiliation{Department of Physics, The Chinese University of Hong Kong, Shatin, N.T., Hong Kong SAR, China}	
\affiliation{Department of Physics and Astronomy, University of Kentucky, 505 Rose St., Lexington, KY 40506-0055, USA}	

\author{J. R. Brownstein}
\affiliation{Department of Physics and Astronomy, University of Utah, 115 S. 1400 E., Salt Lake City, UT 84112, USA}
	
\author{Niv Drory}
\affiliation{McDonald Observatory, The University of Texas at Austin, 1 University Station, Austin, TX 78712, USA}	

\author{Richard R. Lane}
\affiliation{Centro de Investigaci\'on en Astronom\'ia, Universidad Bernardo O Higgins, Avenida Viel 1497, Santiago, Chile}



\begin{abstract}

We combine an unprecedented MaNGA sample of over 3,000 passive galaxies in the stellar mass range $10^{9}-10^{12}$ M$_{\odot}$ with the Sloan Digital Sky Survey group catalog by Tinker to quantify how central and satellite formation, quantified by radial profiles in stellar age, [Fe/H], and [Mg/Fe], depends on the stellar mass of the galaxy (M$_*$) and the mass of the host halo (M$_h$). After controlling for M$_*$ and M$_h$, the stacked spectra of centrals and satellites beyond the effective radius ($r_e$) show small, yet significant differences in multiple spectral features at the 1\% level. According to spectral fitting with the code \texttt{alf}, a primary driver of these differences appears to be [Mg/Fe] variations, suggesting that stellar populations in the outskirts of satellites formed more rapidly than the outer populations of centrals. To probe the physical mechanisms that may be responsible for this signal, we examined how satellite stellar populations depend on M$_h$. We find that satellites in high-M$_h$ halos show older stellar ages, lower [Fe/H], and higher [Mg/Fe] compared to satellites in low-M$_h$ halos, especially for M$_*=10^{9.5}-10^{10.5}$ M$_{\odot}$. These signals lend support to environmentally driven processes that quench satellite galaxies, although variations in the merger histories of central and satellite galaxies also emerge as a viable explanation.

\end{abstract}



\section{Introduction}
\label{1}

Evolution of structure in the $\Lambda$CDM model is hierarchical (\citealt{white-rees1978,davis1985}). Massive central galaxies are thought to grow in stellar mass (M$_*$) and size through the accretion of stellar envelopes from satellite galaxies (\citealt{oser2010,johansson2012,oser2012}). Several observations have found supporting evidence for this picture. Cluster galaxies show an excess in surface brightness that can extend out to 100 kpc and beyond (intra-cluster light; \citealt{zibetti2005}). Similarly, the surface brightness profiles of massive elliptical galaxies (M$_*>10^{11}$ M$_{\odot}$) contain faint, extended components at large radii ($r>10$ kpc; \citealt{huang2013a,huang2013b}) presumably accreted from satellite galaxies (\citealt{huang2018}). Furthermore, in \citet{oyarzun2019} we showed that the stellar metallicity profiles of M$_*>10^{11}$ M$_{\odot}$ early-type galaxies (ETGs) flatten beyond the effective radius ($r_e$), which is another signature of stellar accretion (\citealt{cook2016,taylor2017}).

The hierarchical formation scenario can also successfully explain some of the differences  between the central and satellite galaxy populations. At fixed M$_*$, satellite galaxies show higher stellar concentrations, older stellar populations, higher stellar metallicities, and a higher quenched fraction than central galaxies (\citealt{vandenbosch2008,pasquali2010,wetzel2012,labarbera2014,davies2019,pasquali2019,gallazzi2020,trussler2021}). These observations suggest that parent halos facilitate satellite quenching through mechanisms that can, for example, remove the subhalo ISM (i.e. ram-pressure stripping; \citealt{gunn-gott1972,einasto1974,nulsen1982}) or inhibit further star formation through the shutdown of cold gas accretion (e.g. starvation; \citealt{kawata-mulchaey2008}).

At the low-M$_*$ end (M$_*<10^{8}$ M$_{\odot}$), ram-pressure stripping is thought to dominate satellite quenching (\citealt{weisz2015,davies2016}). Low-M$_*$ satellites interact with the intra-cluster medium (ICM) of the parent halo, stripping their gas reservoirs and truncating their star-formation (e.g. \citealt{balogh-morris2000,mayer2001,mayer2006,vandenbosch2008b,spindler-wake2017}). As a result, satellite galaxies show older stellar populations than centrals of the same M$_*$ (\citealt{pasquali2010}). At the same time, low-M$_*$ satellites can lose some of their M$_*$ through tidal stripping (\citealt{kang2008}). As this process leaves stellar metallicity roughly unaltered (\citealt{pasquali2015}), low M$_*$ satellites deviate from the average stellar mass-metallicity relation followed by central galaxies (\citealt{faber-jackson1976,cidfernandes2005,gallazzi2005,thomas2005,thomas2010,gonzalez-delgado2014,pasquali2015}). 

The impact of ram-pressure stripping is believed to decrease toward higher stellar masses (M$_*=10^{8}-10^{11}$ M$_{\odot}$), as deeper potentials can more effectively sustain drag from the ICM (\citealt{fillingham2015}). At these masses, mechanisms that inhibit future star formation through environmental preprocessing are more likely to dominate. In starvation, satellite galaxies can lose their cold gas reservoirs through tidal interactions with the parent halo, suppressing future star-formation (\citealt{kawata-mulchaey2008}). This process can also apply to hot subhalo gas in a process known as strangulation (e.g. \citealt{larson1980}). These two mechanisms are thought to act in timescales of 2-6 Gyr, after which the satellite galaxy quenches within 1 Gyr (\textsl{delayed-then-rapid}; \citealt{wetzel2012,wetzel2013}).

All of these mechanisms point to a strong connection between the satellite infall time (T$_{inf}$) and the satellite quenching time (T$_{q}$). To test this connection, \citet{pasquali2019} and \citet{smith2019} parameterized T$_{inf}$ in projected phase space (distance and velocity of the satellite within the parent halo). Using this parameterization, \citet{gallazzi2020} found ancient infallers (T$_{inf}>5$ Gyr) to host older stellar populations and show higher stellar metallicities than recent infallers (T$_{inf}<2$ Gyr). This result led \citet{gallazzi2020} to conclude that ancient infallers likely quench through starvation and strangulation (T$_{q}>$T$_{inf}$). On the other hand, recent infallers probably quenched through internal processes, long before they entered the virial radius of the parent halo (T$_{q}<$T$_{inf}$).

Although this picture appears compelling, the true origin of some of the observational differences between centrals and satellites has been questioned. \citet{wang2018a} showed that the quenched fraction, star-formation rates, and 4000\AA\ breaks of centrals and satellites in the Sloan Digital Sky Survey (SDSS) are indistinguishable after both stellar mass and halo mass are controlled for. Similar results were later found for other galaxy properties like size and bulge-to-total light ratio (\citealt{wang2020}). The similarities between the scaling relations followed by centrals and satellites could indicate that their quenching processes are not as different as initially thought (\citealt{wang2018c}).

At this juncture, discerning whether satellite galaxies are subject to unique environment-driven processes can greatly benefit from spatially resolved spectroscopy. For instance, characterization of the stellar mass surface density profiles of satellites can reveal evidence of tidal stripping beyond the tidal radius (\citealt{read2006}). Alternatively, modeling of the star-formation histories of satellite outskirts could reveal signatures of starvation or gas stripping (e.g. \citealt{cortese2021}). These arguments have motivated several works to search for differences between the stellar population gradients of central and satellite galaxies. So far, all analyses with the MaNGA (\citealt{bundy2015}) and SAMI (\citealt{allen2015}) surveys have found no significant differences between the stellar age, metallicity, and [$\alpha$/Fe] gradients of central and satellite galaxies (\citealt{goddard2017b,zheng2017,santucci2020}).

The more direct interpretation of these results is that central and satellites quench in a similar fashion, in alignment with the interpretation by \citet{wang2018a,wang2018c} and in contrast with the \textsl{delayed-then-rapid} scenario. Alternatively, differences between the spatially resolved stellar populations of centrals and satellites could also be very subtle, requiring larger samples to isolate any signatures. For instance, \citet{greene2015} were unable to compare the stellar population profiles of central and satellite galaxies at fixed M$_*$ due to the size of the MASSIVE sample ($\sim 100$ galaxies). Less than 1,000 galaxies had been observed by MaNGA at the time that \citet{goddard2017b} and \citet{zheng2017} performed their work on the stellar population gradients. This number is in strong contrast to the SDSS samples used by \citet{pasquali2010,gallazzi2020,trussler2021}, which exceeded 500,000 galaxies. However, now that the MaNGA survey is complete, in this paper we take advantage of the full sample that exceeds 10,000 galaxies. Taking into account signal to noise and the number of spectra per galaxy, the constraining power of MaNGA is now almost a factor of 2 larger than that of the SDSS \textsc{main} Galaxy Survey Sample (\citealt{york2000,gunn2006,dr12}; \citealt{oyarzun2022}).

Another possible source of uncertainty comes from stellar population characterization. Most fitting codes do not fit for individual element abundances, adding uncertainties to reported ages and metallicities (\citealt{conroy2013}). This is particularly relevant in light of variations in the abundance pattern of passive galaxies with the local environment (e.g. \citealt{greene2019,oyarzun2022}). To account for these variations, in this work we use the stellar population fitting code \texttt{alf} (\citealt{conroy2012a,conroy2018}). This program is designed to capture uncertainties in stellar evolution, allowing us to fit for the age, abundance of various elements, and initial mass function (IMF) of stellar systems older than 1 Gyr.

The use of stellar population gradients in spatially resolved surveys adds to the difficulties. Linear fits to the stellar population profiles of ETGs can \textsl{wash out} underlying trends in the data (\citealt{oyarzun2019}). For example, \citet{greene2015} were only able to conclude that the ages of ETGs correlate with group richness by directly analyzing the stellar age profiles and their dependence on galaxy number density. In this paper, we overcome the limitations posed by the use of gradients by comparing the full extent of the stellar population profiles of central and satellite galaxies.

Cross contamination of central and satellite galaxy samples also presents a challenge. Group catalogs can sometimes struggle to reproduce the fractions of red and blue satellites, biasing the samples (\citealt{tinker2020a}). To mitigate this effect, we employ the SDSS group catalog by \citet{tinker2020b}, which is calibrated on observations of color-dependent galaxy clustering and estimates of the total satellite luminosity to accurately reproduce the fraction of red and blue satellites for M$_*>10^{11}$ M$_{\odot}$. Furthermore, the \citet{tinker2020b} group catalog exploits deep photometry from the DESI Legacy Imaging Survey (\citealt{dey2019}), allowing for the precise measurements of group luminosity required to estimate M$_h$. 

This paper is structured as follows. In Section \ref{2} we define our sample of passive central and satellite galaxies from SDSS-IV MaNGA. In Section \ref{3}, we describe our handling of the spectra and fitting with \texttt{alf}. We show our results in Section \ref{4} and discuss the implications in Section \ref{5}. We summarize in Section \ref{6}. This work adopts a Kroupa IMF (\citealt{kroupa2001}) for estimating stellar masses. We assume H$_0=70$ km s$^{-1}$Mpc$^{-1}$, and all magnitudes are reported in the AB system (\citealt{oke1983}). 
	
\section{Dataset}
\label{2}

\subsection{The MaNGA survey}
\label{2.1}

The MaNGA survey (\citealt{bundy2015,yan2016b}) was part of SDSS-IV (\citealt{york2000,gunn2006,blanton2017,dr15}) and provided spatially resolved spectroscopy for over 10,000 nearby ($z<0.15$) galaxies (\citealt{drory2015,law2015}). The spectra have a median spectral resolution of $\sigma=72$ km s$^{-1}$ ($R\sim2,000$) and cover the wavelength range 3,600-10,300 \AA\ (\citealt{smee2013}). The data cubes typically reach a 10$\sigma$ continuum surface brightness of 23.5 mag arcsec$^{-2}$, and their astrometry is measured to be accurate to 0\farcs1 (\citealt{law2016}). Radial coverage reaches between 1.5 and 2.5 $r_e$ for most targets (\citealt{wake2017}). 

The data was reduced by the MaNGA Data Reduction Pipeline (DRP; \citealt{yan2016a,law2016}). De-projected distances, stellar kinematic maps, and emission line fluxes were computed by the MaNGA Data Analysis Pipeline (DAP; \citealt{belfiore2019,westfall2019}). Effective radii ($r_e$) for all MaNGA galaxies were retrieved from the NASA-Sloan Atlas\footnote{\href{http://nsatlas.org}{http://nsatlas.org}}(NSA). These $r_e$ were determined using an elliptical Petrosian analysis of the $r$-band image from the NSA, implementing the detection and deblending technique described in \citet{blanton2011}. For data access and handling, we used the tool Marvin\footnote{\href{https://www.sdss.org/dr17/manga/marvin/}{https://www.sdss.org/dr17/manga/marvin/}} (\citealt{cherinka2019}).

\begin{figure*}
\centering
\includegraphics[width=7.1in]{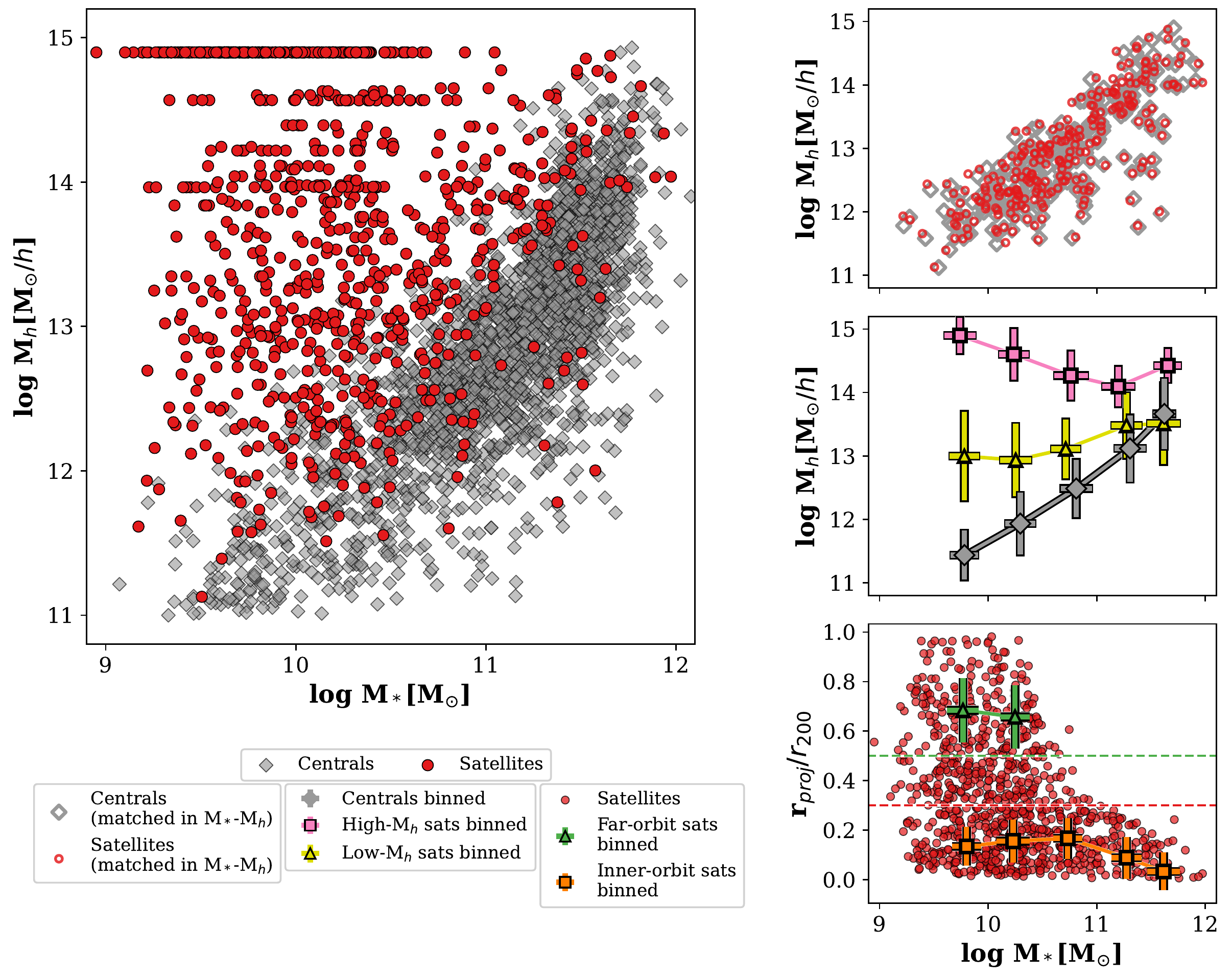}
\caption{Distribution of passive MaNGA galaxies in stellar mass, halo mass, and normalized cluster-centric distance. Central/satellite classification and halo mass estimates come from the \citet{tinker2020b} catalog. Left: stellar-to-halo mass relation for centrals (gray) and satellites (red). On the right-hand side, the different panels show the different galaxy subsamples adopted in our analysis. Top right: central (gray) and satellite (red) subsamples with matching M$_*$ and M$_h$. Middle right: central and satellite subsamples selected according to host halo mass. The average M$_*$ and M$_h$ of high-M$_h$ and low-M$_h$ satellites are shown as magenta squares and yellow triangles, respectively. The error bars show the standard deviation around the average values. Bottom right: satellite subsamples selected on normalized cluster-centric distance. The M$_*$ and cluster-centric distances of inner-orbit satellites and far-orbit satellites are plotted in orange and green, respectively.}
\label{fig0}
\end{figure*}

\begin{deluxetable*}{|c|c|c|c|c|c|c|c|}
	\centering
	\tablecaption{Number of galaxies in each subsample as a function of M$_*$\tablenotemark{a}}
	\tabletypesize{\footnotesize}
	\tablewidth{0pt}
	\tablehead{
		\colhead{} & \colhead{M(M$_{\odot}) = 10^{9.5}-10^{10}$} & \colhead{$10^{10}-10^{10.5}$}  & \colhead{$10^{10.5}-10^{11}$} & \colhead{$10^{11}-10^{11.5}$} & \colhead{$10^{11.5}-10^{12}$} & \colhead{Out of Range} & \colhead{Total}}
	\startdata
	\label{table1}
	Centrals & 111 & 211 & 514 & 948 & 412 & 21 & 2217\\
	Satellites & 273 & 319 & 133 & 67 & 25 & 85 & 902\\
	\hline	
	Cens. (M$_*$-M$_h$ matched) & 27 & 69 & 64 & 54 & 23 & 4 & 241\\
	Sats. (M$_*$-M$_h$ matched) & 27 & 70 & 63 & 54 & 23 & 4 & 241\\
	\hline	
	Low-M$_h$ sats. & 132 & 154 & 88 & 26 & 12 & 38 & 450 \\	
	High-M$_h$ sats. & 141 & 165 & 45 & 41 & 13 & 47 & 452 \\	
	\hline	
	Inner-orbit sats. & 112 & 160 & 109 & 67 & 25 & 31 & 504 \\	
	Far-orbit sats. & 90 & 90 & 5 & 0 & 0 & 35 & 220 \\	
	\hline	
	Low-M$_h$ inner sats. & 61 & 75 & 71 & 26 & 12 & 14 & 259\\	
	High-M$_h$ inner sats. & 51 & 85 & 38 & 41 & 13 & 17 & 245\\	
	\hline	
	Low-M$_h$ far sats. & 46 & 45 & 4 & 0 & 0 & 20 & 115\\	
	High-M$_h$ far sats. & 44 & 45 & 1 & 0 & 0 & 15 & 105\\	
	\enddata
\tablenotetext{a}{Note that not all satellite galaxies are classified as either inner orbit or far orbit (bottom right panel of Figure \ref{fig0}).}
\end{deluxetable*}		

\subsection{Sample}
\label{2.2}

This work uses the final internal release of MaNGA data, known as MaNGA Product Launch 11 (MPL-11). The total number of galaxies in MPL-11 is 10,086. The approach to stellar population characterization that we implement in this paper is designed for old stellar systems only. Therefore, we removed star-forming systems by setting the criterion $\log{(\mbox{sSFR})}<-11.5$ M$_{\odot} yr^{-1}$, where sSFR stands for spatially integrated specific star-formation rates measured in MaNGA as part of the pipeline for the Pipe3D Value Added Catalog for DR17\footnote{\href{https://www.sdss.org/dr17/manga/manga-data/manga-pipe3d-value-added-catalog/}{https://www.sdss.org/dr17/manga/manga-data/manga-pipe3d-value-added-catalog/}.} (\citealt{sanchez2016a,sanchez2022}) for Data Release 17 (\citealt{dr17}). Since these sSFRs were corrected for dust attenuation using the Balmer decrement (\citealt{sanchez2016b}), this method of estimating the intrinsic sSFR of galaxies is among the most reliable at low redshift (e.g. \citealt{moustakas2006}). This selection resulted in a sample of 3957 passive galaxies, of which 2217 have more than a 90\% probability of being a central and 902 have more than a 90\% probability of being a satellite according to the \citet{tinker2020b} catalog (further details in Section \ref{2.4}).
	
\subsection{Stellar masses}
\label{2.3}

To estimate the stellar mass (M$_*$) of every galaxy, we first co-added the spectra within the $r_e$. Then, the mass within the $r_e$ was measured by running the stellar population fitting code \texttt{Prospector}\footnote{\href{https://github.com/bd-j/prospector/blob/master/doc/index.rst}{Prospector \\ https://github.com/bd-j/prospector/blob/master/doc/index.rst}}(\citealt{leja2017}) on the co-added spectrum.

\texttt{Prospector} samples the posterior distribution for a variety of stellar population parameters and star formation history (SFH) prescriptions. Stellar population synthesis is handled by the code FSPS\footnote{\href{https://github.com/cconroy20/fsps}{FSPS: Flexible Stellar Population Synthesis \\ https://github.com/cconroy20/fsps}}(\citealt{conroy2009,conroy-gunn2010}). Our runs adopted the MILES stellar library (\citealt{sanchez-blazquez2006}), MIST isochrones (\citealt{choi2016,dotter2016}), and a \citealt{kroupa2001} IMF. For the SFH, we implemented a nonparametric prescription with a continuity prior, emphasizing smooth SFHs over time (\citealt{leja2019}).

The total stellar mass of the galaxy is then
\begin{gather}
\mbox{M}_*^{total}= 2\mbox{M}_*^{r_{\mbox{\scriptsize e}}} \times 10^{-0.15}, 
\end{gather}

where $\mbox{M}_*^{r_{\mbox{\scriptsize e}}}$ is the spectroscopic stellar mass within the effective radius. We measure an offset of 0.15 dex between our $2\mbox{M}_*^{r_{\mbox{\scriptsize e}}}$ and the stellar masses measured through \textsl{k}-correction fits to the Sersic fluxes in the NSA (\citealt{blanton2007}). This is not unexpected, since half-mass radii are smaller than half-light radii (\citealt{garcia-benito2017}). We correct for this offset by multiplying our stellar masses by $10^{-0.15}$ (see the equation). In the rest of the paper, we will simply refer to M$_*^{total}$ as M$_*$.

\subsection{Local environment}
\label{2.4}

This paper uses the \citet{tinker2020b} group catalog for the characterization of the local environment. This catalog is the implementation on SDSS of the self-calibrating halo-based galaxy group finder (\citealt{tinker2020a}). In the finder algorithm, the probability of a galaxy being a satellite is dependent on both galaxy color and luminosity. This allows \citet{tinker2020b} to accurately reproduce the fraction of massive quenched satellite galaxies and estimate M$_h$ more accurately (\citealt{campbell2015}). The \citet{tinker2020b} catalog also took advantage of deep photometry from the DESI Legacy Imaging Survey (\citealt{dey2019}), allowing for precise group and galaxy M$_*$ measurements. These improvements are key to accurately constraining M$_h$ (\citealt{bernardi2013,wechsler2018}).

Yet, the approach in \citet{tinker2020a} still has its limitations. Like most group catalogs, it is susceptible to central galaxy misidentification. It also assumes that the total satellite luminosity is a function of halo mass only. The \citet{tinker2020b} catalog also fails to reproduce the clustering of faint quiescent galaxies (M$_*<10^{9.5}$ M$_{\odot}$). We minimize the impact of these biases by setting strict central and satellite identification criteria (next paragraph) and by avoiding the lowest M$_*$ end of the galaxy population (our galaxies have M$_*>10^{9}$M$_{\odot}$; Figure \ref{fig0}).

This work used the public version of the \citet{tinker2020b} group catalog\footnote{\href{https://galaxygroupfinder.net}{https://galaxygroupfinder.net}}. Satellite probabilities (P$_{sat}$) were used to define central and satellite galaxy subsamples. To select centrals, we implemented the criterion P$_{sat}<0.1$. To select satellites, we set P$_{sat}>0.9$. These selections yielded two subsamples with 2217 central and 902 satellite galaxies. Their distribution in M$_h$ and M$_*$ space is shown in Figure \ref{fig0}.

\section{Methodology}
\label{3}

\subsection{Central and satellite subsamples}
\label{3.1}

This section presents the different subsamples of central and satellite galaxies. The overall distribution of our centrals and satellites in M$_*$ and M$_h$ is plotted in the left panel of Figure \ref{fig0}. To properly compare the samples, we constructed 241 pairs of centrals and satellites that have matching M$_*$ and M$_h$ (top right panel of Figure \ref{fig0}). To characterize how satellite formation depends on M$_h$, we defined two more satellite subsamples: low-M$_h$ satellites and high-M$_h$ satellites (middle right panel of Figure \ref{fig0}). These subsamples were defined to include only satellites below and above the 50th percentile in M$_*$-to-M$_h$ ratio as a function of M$_*$. We note that the difference in M$_h$ between the two satellite subsamples decreases with M$_*$.

The epoch at which the satellite first crosses the virial radius of the host halo (infall time; T$_{inf}$) is another property that could shape the stellar populations of satellite galaxies. \citet{gallazzi2020} selected ancient (T$_{inf}>5$ Gyr) and recent (T$_{inf}<2.5$ Gyr) infallers by characterizing how T$_{inf}$ is mapped onto the projected phase space composed of cluster-centric velocity and distance (\citealt{pasquali2019,smith2019}). Here, we implement a similar, albeit more simplistic, analysis based on cluster-centric distance only. As in \citet{pasquali2019} and \citet{smith2019}, we define the projected, normalized cluster-centric distance as
\begin{gather}
D_{proj}= r_{proj}/r_{200} \\
r_{200}[\mbox{kpc h}^{-1}]= 258.1 \times \frac{(M_h/10^{12})^{1/3} \times (\Omega_m/0.25)^{1/3}}{(1+z)} \mbox{ ,} \nonumber
\end{gather}

where $r_{proj}$ is the projected distance between the central and satellite, $r_{200}$ is the virial radius (\citealt{yang2007,pasquali2019}), $\Omega_m=0.3$, and $h=0.7$.

Based on $D_{proj}$, we defined two more subsamples. inner-orbit satellites have $D_{proj}<0.3$, and are meant to resemble the ancient infaller selection. According to Table 1 in \citet{pasquali2019}, inner-orbit satellites have an average $\bar{\mbox{T}}_{inf}\sim 5$ Gyr and standard deviation $\sigma_{inf}\sim 2.5$ Gyr. Based on these numbers, we estimate a recent infaller contamination of 15\%.

On the other hand, far-orbit satellites are defined by $D_{proj}>0.5$ and are similar to recent infallers. Based on Table 1 from \citet{pasquali2019}, far-orbit satellites have $\bar{\mbox{T}}_{inf}\sim 3.5$ Gyr and $\sigma_{inf}\sim 2.5$. We estimate an ancient infaller contamination of around 20\%. The subsamples obtained in this cluster-centric distance selection are plotted in the bottom right panel of Figure \ref{fig0}. 

Detailed number counts for all subsamples are presented in Table \ref{table1}. Note that not all satellite galaxies are classified as either inner orbit or far orbit. After defining the subsamples, we computed high signal-to-noise stacked spectra through the process described in the following section.

\subsection{Co-addition and stacking of spectra}
\label{3.2}

Within every galaxy, we co-added the spectra into the three annuli $r= [0, 0.5]$, $[0.5, 1]$, and $[1, 1.5]$ effective radii. This step required shifting every spectrum back to the rest frame using the stellar systemic velocity ($v_*$) maps calculated by the DAP. Then, we stacked the spectra of central and satellite galaxies in each M$_*$ and annular bin. Stacks were obtained by computing the median and errors were quantified through Monte Carlo simulations that accounted for the propagated errors. All spectra were convolved to $\sigma_*=350\ km s^{-1}$ and median normalized before stacking. 
Greater detail on the stacking process can be found in \citet{oyarzun2022}. 

\begin{figure*}
	\centering
	\includegraphics[width=7.1in]{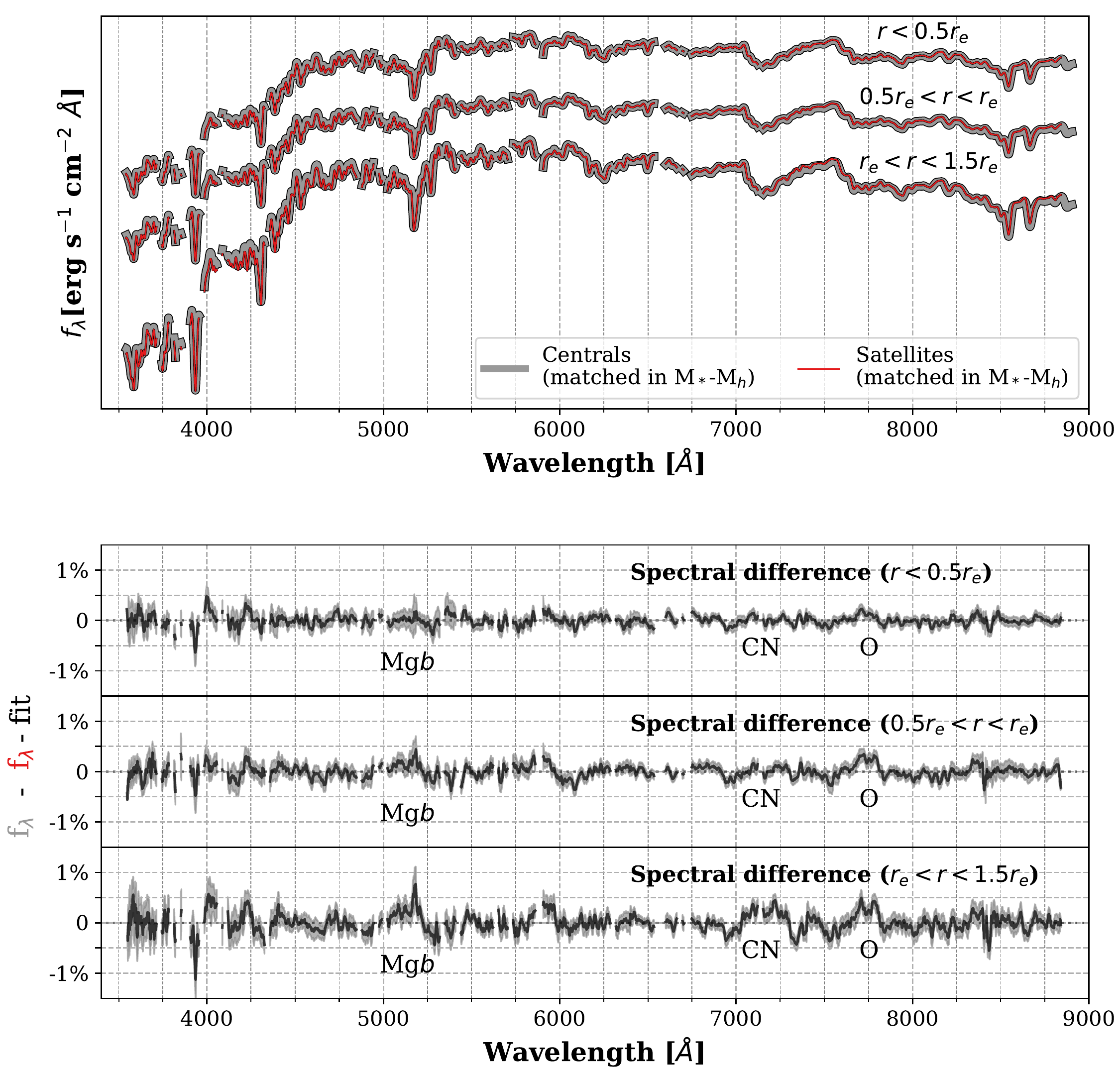}
	\caption{Direct spectral comparison between central and satellite galaxies with matching M$_*$ and M$_h$. Top: stacked spectra for centrals (gray) and satellites (red) in three radial bins extending from the centers ($r<0.5 r_e$; top) to the outskirts ($r_e<r<1.5 r_e$; bottom). Bottom: spectral difference obtained by subtracting the satellite stack from the central stack in each radial bin. A polynomial was fit to the spectral difference to remove continuum shape variations. Differences are shown in percent relative to the median of the spectrum. The gray shading shows the error on the differences. Centrals and satellites show several significant spectral differences at the 1\% level that increase in significance and magnitude as galactocentric distance increases.}
	\label{fig1}
\end{figure*}	

\begin{figure*}
	\centering
	\includegraphics[width=7.1in]{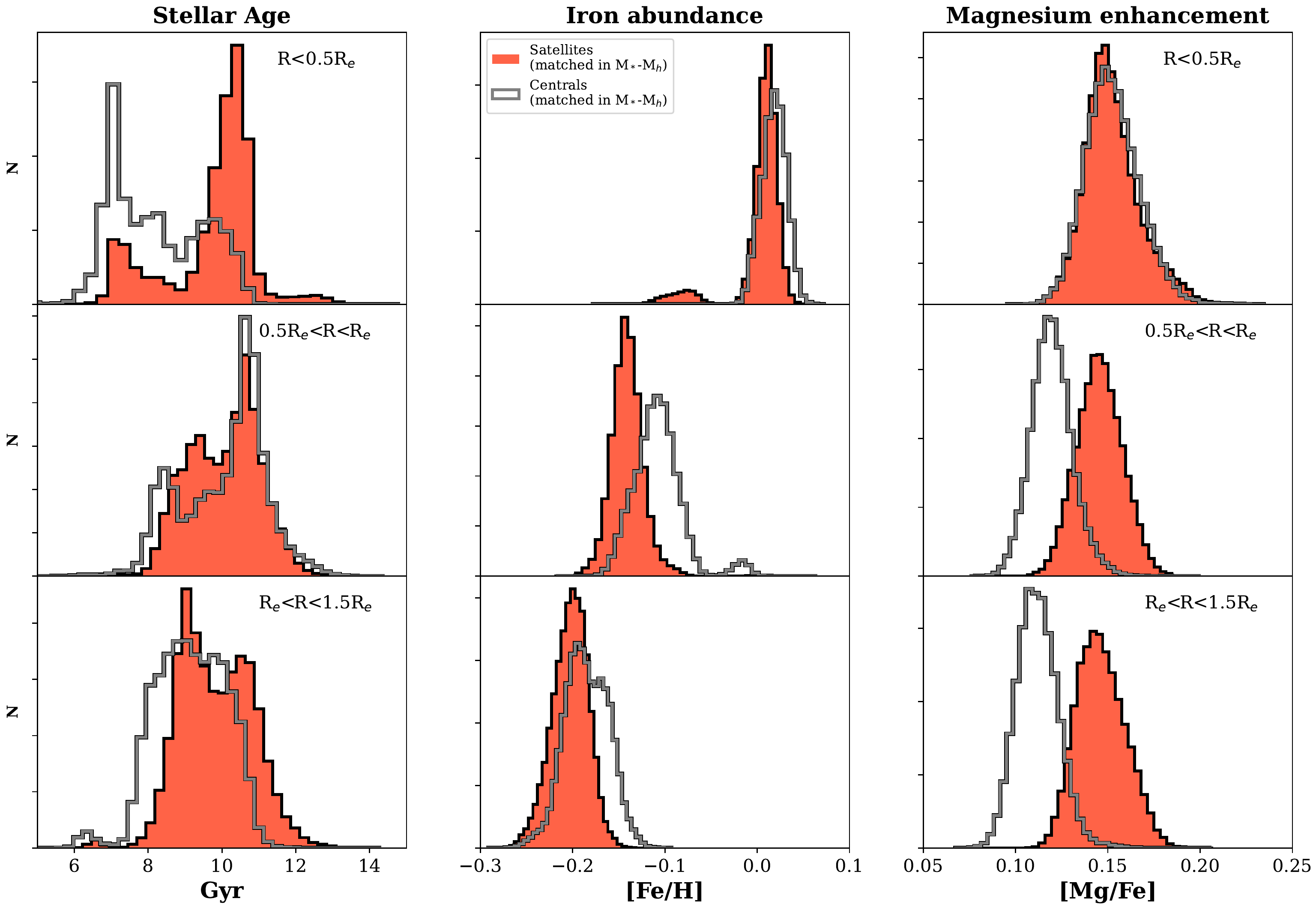}
	\caption{Posterior distributions derived with \texttt{alf} for the stellar age, [Fe/H], and [Mg/Fe] of centrals and satellites with matching M$_*$ and M$_h$ distributions (Figure \ref{fig1}). Central galaxy posteriors are shown in gray, whereas satellite galaxy posteriors are shown in red. Galactocentric distance increases from top to bottom. Satellites are older than centrals in the centers and more Mg-enhanced in the outskirts.} 
	\label{fig2}
\end{figure*}	

\begin{figure*}
	\centering
	\includegraphics[width=7.1in]{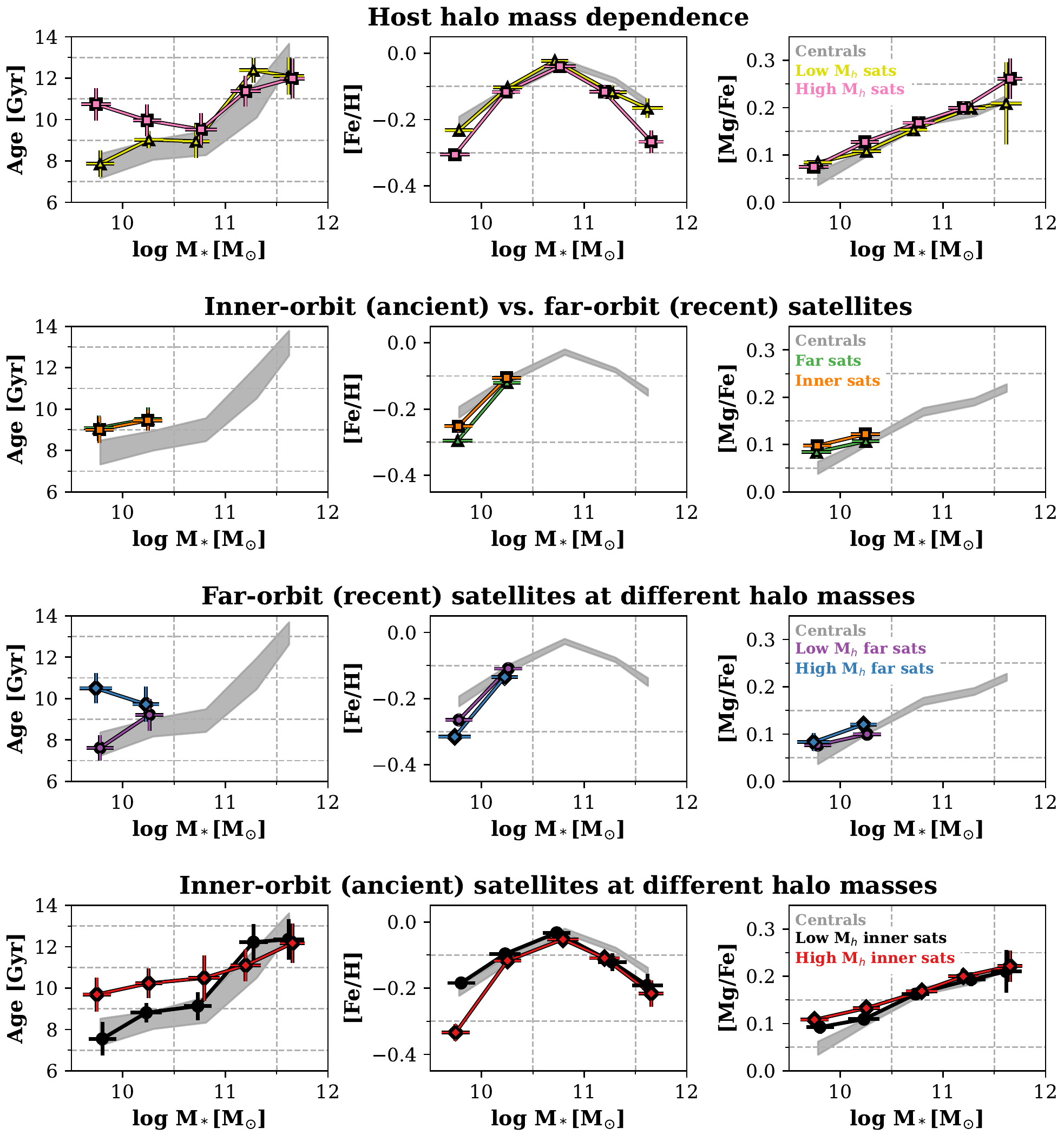}
	\caption{Stellar population parameters integrated within 1.5 $r_e$ as a function of M$_*$. Stellar age is shown on the left, [Fe/H] in the middle, and [Mg/Fe] on the right. Different rows compare centrals against different subsamples of satellite galaxies. In the top two rows, satellites are divided as a function of host halo mass and cluster-centric distance. Both selection methods are combined in the bottom two rows, with far-orbit satellites shown in the third row and inner-orbit satellites at the bottom. For M$_*<10^{10.5}$ M$_{\odot}$, satellites found in high-M$_h$ halos are old, have lower [Fe/H], and show higher [Mg/Fe] than centrals, even at fixed cluster-centric distance.}
	\label{fig3}
\end{figure*}	

\subsection{Stellar population fitting with \texttt{alf}}
\label{3.3}

The stacked spectra were fitted with the code \texttt{alf} to characterize their stellar populations. The program \texttt{alf} fits the optical absorption line spectra of old ($\gtrsim$1 Gyr) stellar systems (\citealt{conroy2012a,conroy2018}) to estimate their stellar population parameters. It is based on the MIST isochrones (\citealt{choi2016,dotter2016}) and the stellar libraries by \citet{sanchez-blazquez2006} and \citet{villaume2017}. Deviations from the solar abundance pattern are quantified in the theoretical response functions (\citealt{conroy2018,kurucz2018}).

We fitted for a two-component SFH (i.e. two SSPs), stellar velocity dispersion, IMF, and the abundances of 19 elements. For the IMF, power laws were fit in the ranges 0.08-0.5 M$_{\odot}$ and 0.5-1 M$_{\odot}$, with the IMF slope set to -2.35 for the 1-100 M$_{\odot}$ range (\citealt{salpeter1955}). We sampled this multivariate posterior with \texttt{emcee} (\citealt{emcee}) using a setup of 1024 walkers, 10$^{4}$ burn-in steps, and 100-step chains.

Stellar ages reported throughout correspond to the mass-weighted age of the two-component SFH. Uncertainties on the fitted parameters account for errors in sample assignment, stacking, and fitting. This was achieved by bootstrapping the selection of galaxies 10 times, and then by stacking and fitting in every iteration. 

\begin{figure*}
	\centering
	\includegraphics[width=7.2in]{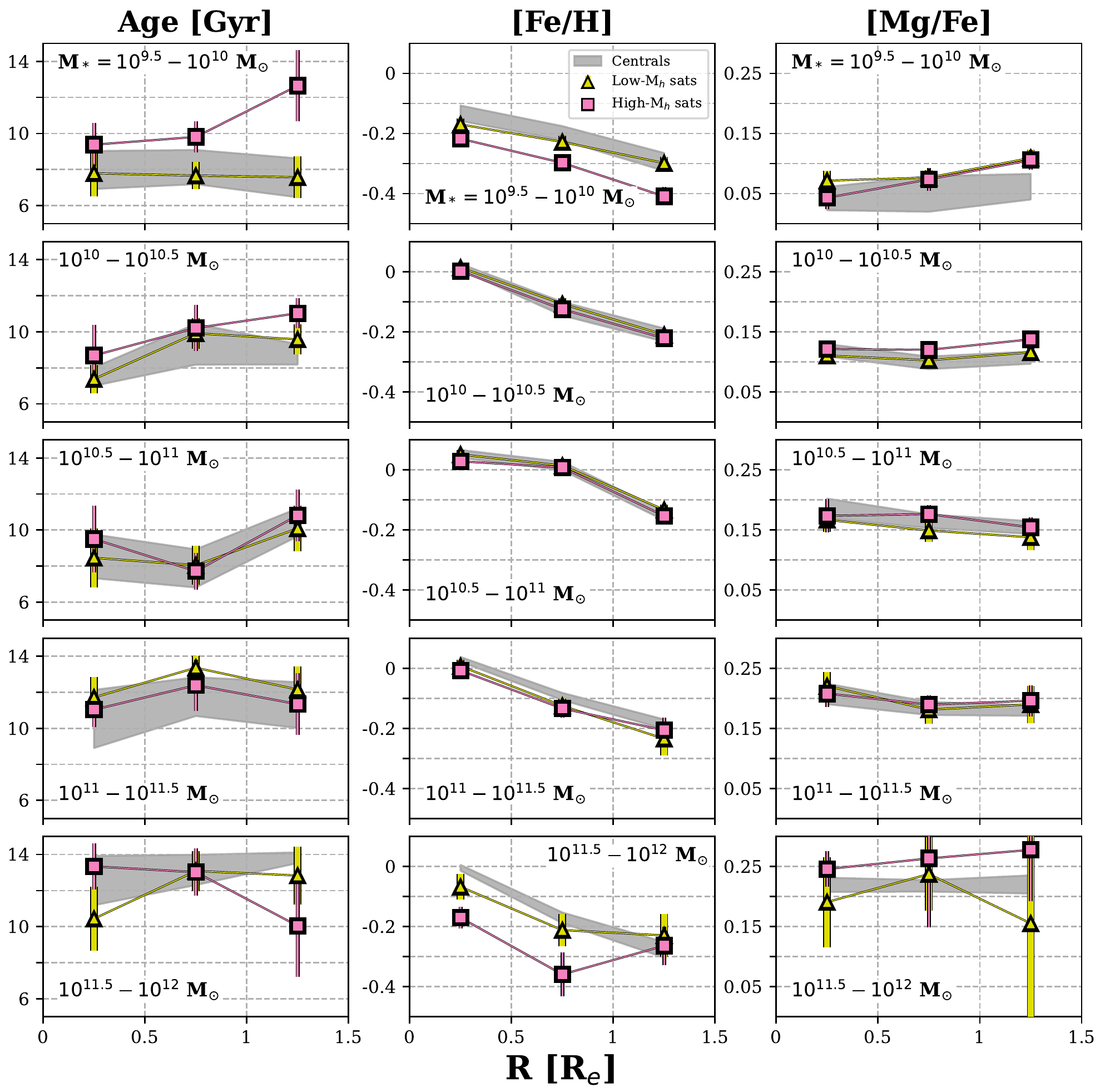}
	\caption{Stellar population profiles of centrals (gray), low-M$_h$ satellites (yellow), and high-M$_h$ satellites (magenta). Stellar mass increases from top to bottom. From left to right, the columns show stellar age, iron abundance, and magnesium enhancement.}
	\label{fig4}
\end{figure*}	

\section{Results}
\label{4}

\subsection{Central and satellite galaxies}
\label{4.1}

Recent works have found that differences between the observable properties of centrals and satellites start to vanish once both M$_*$ and M$_h$ are controlled for (\citealt{bluck2016,wang2018a,wang2018b,wang2018c,wang2020}). We test whether this conclusion is supported by our data in Figure \ref{fig1}. The top panel shows the stacked spectra for a total of 241 centrals and 241 satellites that have matching M$_*$ and M$_h$ (see Figure \ref{fig0}). It becomes apparent that the stacked spectra of centrals and satellites are very similar in shape and in the strength of their absorption features at all radii.

To identify any spectral variations, we plotted the spectral differences between the stacks as a function of galactocentric distance in the second panel of Figure \ref{fig1}. These spectral differences were computed by subtracting the satellite spectrum from the central spectrum and then fitting polynomials to decouple continuum variations. Despite the subtlety of all spectral differences ($\lesssim 1\%$), some show high significance (e.g. Mg$\textsl{b}$ 5172\AA\ ; \citealt{faber-jackson1976,faber1985}). There is also clear indication that the magnitude of the spectral differences increases with galactocentric radius.

To translate these differences into stellar population variations, we fitted the stacked spectra with \texttt{alf}. We derived posteriors for the mass-weighted stellar age, [Fe/H], and [Mg/Fe] in three radial bins extending from centers out to 1.5 $r_e$ (Figure \ref{fig2}). At the centers, satellites are older than centrals by $\sim 2$ Gyr. In the outskirts ($r=0.5$-1.5 $r_e$), satellites show higher [Mg/Fe] by $\lesssim 0.05$ dex.

\subsection{How the stellar populations of satellites depend on M$_h$ and cluster-centric distance}

We now turn to our analysis of the stellar population parameters of centrals and satellites selected as a function of host M$_h$ and cluster-centric distance (Figure \ref{fig0}). Stacked spectra for galaxies within these subsamples were derived and fitted with \texttt{alf}. The recovered stellar population parameters within 1.5 $r_e$ as a function of M$_*$ are presented in Figure \ref{fig3}. These radially integrated values were derived via Monte Carlo simulations of radial averaging from the three radially dependent posterior distributions measured with \texttt{alf}. The datapoints and error bars shown in Figure \ref{fig3} therefore correspond to the median and the standard deviation of the radial-average distributions for every subsample and M$_*$ bin. Instead of running \texttt{alf} on spectra stacked out to 1.5 $r_e$, our approach ensures that the radial averages are not dependent surface brightness, i.e., all radial scales are weighted equally.

Generally, the ages of central and satellite galaxies increase from $\sim$8 Gyr at M$_*=10^{10}$ M$_{\odot}$ to $>10$ Gyr for M$_*>10^{11}$ M$_{\odot}$, in agreement with previous observations that galaxies of increasing M$_*$ are older (e.g. \citealt{gallazzi2005,mcdermid2015,lacerna2020}). As for [Fe/H], its magnitude increases with M$_*$ before turning over around M$_*\sim 10^{11}$ M$_{\odot}$. This might seem to contradict the stellar mass-metallicity relation, in which metallicity always increases with the M$_*$ or central velocity dispersion of the galaxy (\citealt{faber-jackson1976,cidfernandes2005,gallazzi2005,thomas2005,thomas2010,gonzalez-delgado2014,mcdermid2015}). However, stellar metallicity typically refers to a weighted average of the abundance of various elements (i.e., a rescaling of the solar abundance pattern), whereas the [Fe/H] that we measured with \texttt{alf} maps the abundance of iron only. As we pointed out in \citet{oyarzun2022}, the decrease in [Fe/H] with M$_*$ at the massive end is likely a consequence of how the star-formation timescales of galaxies shorten as M$_*$ increases. The abundance of magnesium---a proxy for [$\alpha$/Fe] (e.g. \citealt{faber-jackson1976,faber1985,thomas2005})---monotonically increases with M$_*$, supporting this interpretation.

The first row of Figure \ref{fig3} highlights a comparison between satellites in halos of different masses. Satellites in high-M$_h$ halos are older than counterparts in low-M$_h$ halos by $\sim 2$ Gyr for M$_*<10^{10.5}$ M$_{\odot}$. We also observe slightly lower [Fe/H] and higher [Mg/Fe] in high-M$_h$ satellites.

To quantify the significance of these subtle stellar population differences, we can turn to a Bayesian model comparison. We will consider two models: in model $\mathcal{A}$ the magnitude of the stellar population parameter $\mathcal{X}$ is greater in high-M$_h$ satellites, whereas in model $\mathcal{B}$ the magnitude of $\mathcal{X}$ is greater in low-M$_h$ satellites. By assuming that the two models account for all possible options (e.g. \citealt{oyarzun2017,oyarzun2022}), the probabilities of models $\mathcal{A}$ and $\mathcal{B}$ become
\begin{gather}
\mbox{I\kern-0.15em P($\mathcal{A}$)}= \frac{p_{\mathcal{A}}}{p_{\mathcal{A}}+p_{\mathcal{B}}} \ \wedge \ 
\mbox{I\kern-0.15em P($\mathcal{B}$)}= \frac{p_{\mathcal{B}}}{p_{\mathcal{A}}+p_{\mathcal{B}}} 
\\
\nonumber
\\
p_{\mathcal{A}} =\prod_{\scriptsize \mbox{M}_*} \mbox{I\kern-0.15em P($\mathcal{X}$}_{\mbox{high-M}h} > \mbox{$\mathcal{X}$}_{\mbox{low-M}h})
\\
p_{\mathcal{B}} = \prod_{\scriptsize \mbox{M}_*} \mbox{I\kern-0.15em P($\mathcal{X}$}_{\mbox{low-M}h} < \mbox{$\mathcal{X}$}_{\mbox{high-M}h})
\end{gather}

Integrated over the range M$_*=10^{9.5}-10^{10.5}$ M$_{\odot}$, high-M$_h$ satellites have older ages, lower [Fe/H], and higher [Mg/Fe] than low-M$_h$ satellites of the same M$_*$ with 3.2$\sigma$, 5.7$\sigma$, and 2.4$\sigma$ significance, respectively. Taken together, these results indicate that satellites found in more massive dark matter halos formed their stars, on average, earlier and in shorter timescales.

The stellar populations of satellite galaxies can also depend on properties other than host M$_h$. \citet{gallazzi2020} argued that the old ages of high-M$_h$ satellites can be driven by ancient infallers (T$_{inf}>5$ Gyr), which are primarily found at small cluster-centric distances. We examine how the satellite stellar populations depend on cluster-centric distance in the second row of Figure \ref{fig3}. Although there is no evidence of stellar age differences with cluster-centric distance, there is some evidence that inner-orbit satellites have higher [Fe/H] (2.6$\sigma$) and higher [Mg/Fe] (1.8$\sigma$) than far-orbit satellites.

We also show in Figure \ref{fig3} that the differences between satellites with varying host M$_h$ are also present when cluster-centric distance is controlled for. As can be seen in the third row of this figure, far-orbit satellites in high-M$_h$ halos feature older ages, lower [Fe/H], and higher [Mg/Fe] than far-orbit satellites in low-M$_h$ halos, at least for the M$_*$ range probed by our far-orbit satellite subsample (see Figure \ref{fig0}). The direction and magnitude of these differences with host M$_h$ are also seen in the inner-orbit satellite subsample (fourth row).

Central galaxies are plotted with gray shading to provide a rough point of reference.  A quantitative comparison between centrals and satellites is not possible in Figure 4, however, because these two samples have not been matched in M$_*$ and M$_h$ (for that comparison, see Figure \ref{fig2}). That said, the differences we detected in Figure \ref{fig2} are qualitatively apparent in Figure \ref{fig3}, with all satellite subsamples showing higher [Mg/Fe] in most M$_*$ bins. In regards to how the stellar populations of centrals vary with M$_h$ at fixed M$_*$, we showed in \citet{oyarzun2022} that those variations are small and within the uncertainties shown in Figure \ref{fig3}.

\subsection{The stellar population profiles}

We can also gain insight into how central and satellite galaxies form through the radial distribution of their stellar populations. The profiles at different M$_*$ are compared in Figure \ref{fig4}. From left to right, shown are stellar age, [Fe/H], and [Mg/Fe]. The stellar age profiles are mostly flat within our error bars for all subsamples, in agreement with other work on the stellar population gradients of massive galaxies (\citealt{goddard2017a,goddard2017b,zheng2017,lacerna2020}). On the other hand, the [Fe/H] profiles fall with galactocentric distance (e.g. \citealt{greene2015,parikh2018,parikh2019}). As measured in \citet{parikh2019}, the [Mg/Fe] profiles are constant with galactocentric distance in most cases.

We notice systematic offsets in the normalization of satellite profiles with different M$_h$ that mirror the results shown in Figure \ref{fig3} and described in the previous section. Figure \ref{fig4} reveals that variations in the integrated properties of satellites with M$_h$ (Figure \ref{fig3}) are for the most part systematic with galactocentric distance. Regarding variations in the \textsl{shape} of the profiles, the increased uncertainties resulting from additional binning in galactocentric distance prevent us from detecting any statistically significant signals.

\begin{figure}
	\centering
	\includegraphics[width=3.2in]{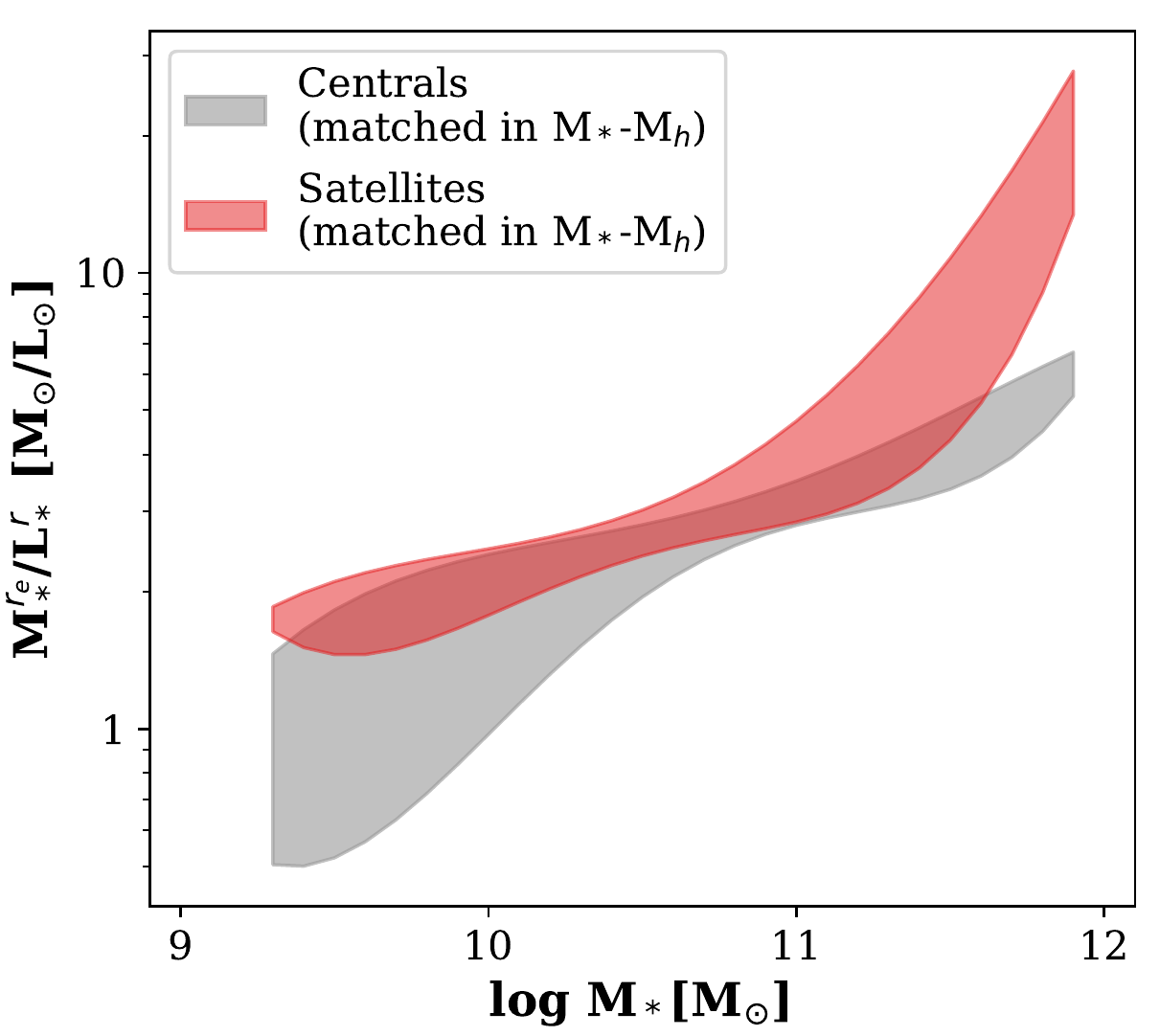}
	\caption{Comparison between the stellar mass-to-light ratios of central and satellite subsamples matched in M$_*$ and M$_h$ (1$\sigma$ contours). Stellar masses were measured in MaNGA and account for the mass within the $r_e$ only. Galaxy luminosities were measured in r-band photometry from the DESI Legacy Imaging Survey (\citealt{dey2019}) as described in \citet{tinker2020b}. Note how centrals have lower M$_*^{r_e}/$L$_*^{r}$, likely as a result of their greater stellar masses beyond the $r_e$ (e.g. \citealt{huang2013a}).}
	\label{fig6}
\end{figure}

\section{Discussion}
\label{5}

We have found that stellar populations in the outskirts of passive satellite galaxies are more alpha enhanced than the stellar populations in the outskirts of passive central galaxies of the same M$_*$ and M$_h$. We also found that passive satellites in high-M$_h$ halos feature older stellar ages, lower [Fe/H], and higher [Mg/Fe] than passive satellites in low-M$_h$ halos, especially for M$_*=10^{9.5}-10^{10.5}$ M$_{\odot}$. We now aim to place these results in a physical context and compare them with previous work.

\subsection{Inferences from direct spectral comparison}

While previous work has found passive satellite galaxies to be older and more alpha enhanced than centrals of the same M$_*$ (e.g. \citealt{gallazzi2020}), it has not been possible to control for a possible bias in host M$_h$ (e.g. \citealt{wang2018a,wang2020}). We know that satellites reside in more massive halos than centrals at fixed M$_*$ (Figure \ref{fig0}) and that centrals become older and more alpha enhanced as M$_h$ increases (\citealt{oyarzun2022}), and thus we might attribute the differences between centrals and satellites to how M$_h$ modulates galaxy formation. 

Figure \ref{fig1} was designed to search for spectral differences between centrals and satellites while controlling for both M$_*$ and M$_h$. We detected significant spectral differences between central and satellite spectra at the 1\% level, which is direct evidence that galaxy formation is not agnostic to whether the galaxy is a central or a satellite. Linking these spectral differences to stellar population variations reveals that satellites are older in the centers and more alpha enhanced in the outskirts than centrals (Figure \ref{fig2}).

Also telling is how the magnitude of these spectral differences increases with galactocentric distance. This could be an indication that the physical processes driving these differences are operating from the outside-in. From a satellite perspective, gas reservoirs in the outskirts may be stripped away by host halo-driven processes like starvation, leading to outside-in quenching (Section \ref{env-driven}). On the other hand, central galaxies may be growing their outskirts through the accretion of smaller satellite galaxies (Section \ref{halo}). Both scenarios are discussed in detail in the following sections.
 
\subsection{The environment-driven quenching scenario}
\label{env-driven}

Through processes like starvation and strangulation, massive halos can strip satellite galaxies of their cold and hot gas reservoirs, suppressing future star-formation (\citealt{larson1980,kawata-mulchaey2008}). These mechanisms are believed to act on timescales of 2-6 Gyr, after which the satellite galaxy would rapidly quench within 1 Gyr, and its stellar populations would then appear old and alpha enhanced (\textsl{delayed-then-rapid}; \citealt{wetzel2012,wetzel2013,fossati2017,cora2019}). This is indeed what we see in Figure \ref{fig2}, and is further apparent among high-M$_h$ halos shown in Figure \ref{fig3}. More massive halos may be exerting stronger tidal forces and thus more efficiently removing gas from satellite galaxies (\citealt{kang2008}). The suppression of star-formation also prevents the satellite galaxy from recycling metals like Fe that form on long timescales. Central galaxies, on the other hand, can keep forming stars for longer, increasing their [Fe/H] relative to satellites (Figures \ref{fig2} and \ref{fig3}). As might be expected with longer timescales of star formation, Figure \ref{fig2} also shows some evidence for younger stellar populations in centrals that have been M$_*$ and M$_h$ matched to satellites.

Using spatially unresolved SDSS observations, \citet{pasquali2010} and \citet{gallazzi2020} found satellite galaxies to show higher stellar metallicities---$\log{Z/Z_{\odot}}$---than centrals of the same M$_*$. The results in \citet{gallazzi2020} are particularly relevant to this work because they were not only recovered in the whole galaxy population, but also in their subsample of passive galaxies. However, it is difficult to directly compare these results with our findings given that $\log{Z/Z_{\odot}}$ depends on the abundance of iron, magnesium, and other alpha elements, which we have shown to vary in different subsamples. Yet, the fact that $\log{Z/Z_{\odot}}$ monotonically increases with M$_*$ has been used to test another environment-driven mechanism: tidal stripping. This mechanism reduces satellite M$_*$ while leaving stellar metallicity unaffected (\citealt{kang2008}). This can cause satellites to show higher metallicities than expected for their M$_*$ (\citealt{pasquali2015}). That said, we note that tidal stripping is thought to dominate at masses lower than the mass range of these works (M$_*<10^{8}$ M$_{\odot}$; \citealt{weisz2015,davies2016,kawinwanichakij2017}).

One implication of environment-driven quenching is that the time of quenching is dependent on the time of infall of the satellite. Starvation and strangulation begin earlier in ancient infallers (T$_{inf} > 5$ Gyr; \citealt{pasquali2019}) compared to recent infallers (T$_{inf} < 2.5$ Gyr; \citealt{pasquali2019}). Ancient infallers also tend to orbit closer to the halo center (\textsl{inner-orbits}), where the external quenching processes are strongest (\citealt{vandevoort2017,cleland2021}). Indeed, \citet{gallazzi2020} found ancient infallers to have older ages, higher $\log{Z/Z_{\odot}}$, and greater [$\alpha$/Fe] than recent infallers for M$_*\sim 10^{10}$ M$_{\odot}$.

However, \citet{gallazzi2020} did not separate satellite galaxies into star-forming and passive for their time-of-infall analysis. In this paper, we studied passive galaxies only and adopted a cruder definition of \textsl{ancient} and \textsl{recent} infallers. Our result is shown in the second row of Figure \ref{fig3}. While far more subtle than in \citet{gallazzi2020}, we see some evidence for increased [Fe/H] and [Mg/Fe] among the inner-orbit satellites that we associate with ancient infallers. 

Focusing on ancient infallers only, \citet{gallazzi2020} found their stellar ages and [$\alpha$/Fe] to correlate with host M$_h$, in agreement with our Figure \ref{fig3} (bottom row). Their analysis of recent infallers, on the other hand, yielded little stellar population variations with host M$_h$. In contrast, other work has found evidence that recent infallers feature lower sSFR and older ages than field galaxies of the same M$_*$ (\citealt{pasquali2019}). In addition, \citet{smith2019} found evidence that clusters impact the star-formation histories of satellite galaxies before they enter the virial radius of the host halo. Our results from Figure \ref{fig3} better align with the findings in \citet{pasquali2019} and \citet{smith2019}, in support of a picture where massive halos can affect the assembly of satellites in far orbits as well.

This notion is often associated with \textsl{preprocessing}. Here, satellite cold gas reservoirs are heated up (i.e., evaporation) and/or warm subhalo gas is stripped (i.e. strangulation) by the hot ICM gas (\citealt{fujita2004}) even before they enter the virial radius of the host halo (\citealt{pallero2019}). This scenario not only can explain the differences with host M$_h$ in the populations of recent infallers, but also subtle variations in the quenched fraction of galaxies at large cluster-centric distances (\citealt{haines2015,bianconi2018,vanderburg2018,sarron2019,sarron2021}).

Cosmological simulations also provide support for this scenario. \citet{lacerna2022} found that quenched central galaxies of M$_*\sim 10^{10}$ M$_{\odot}$ are, on average, 5 times closer to the nearest massive group or cluster than star-forming centrals of the same M$_*$. This result not only is consistent with preprocessing, but can also explain the similarity between the properties of galaxies in adjacent halos (i.e., two-halo conformity; \citealt{kauffmann2013,hearin2015}).

In summary, we have shown that centrals and satellites show differences in their assembly histories that cannot be fully ascribed to M$_h$ variations. Environment-driven quenching provides an explanation for these trends, qualitatively explaining why satellites have older ages, lower [Fe/H], higher [Mg/Fe], and higher quenched fractions (\citealt{davies2019}) than centrals, even after controlling for both M$_*$ and M$_h$. That said, our picture of environment-driven quenching needed to be broadened to include preprocessing effects that occur at very large cluster-centric distances before infalling galaxies enter the halo.

\subsection{Stellar population variations in the outskirts from an environment-driven quenching perspective}

Through processes like gas ram pressure and evaporation, host halos can strip weakly bound neutral gas from the outskirts of satellite galaxies via tidal and hydrodynamical effects, inhibiting the formation of \HII\ in the outer disk (\citealt{cortese2021}) and quenching the galaxy from the outside-in (\citealt{chung2009}). This has motivated a search for variations in the stellar population parameters of satellites with galactocentric distance, although no conclusive evidence has been found (e.g. \citealt{goddard2017b,zheng2017,santucci2020,zhou2020}).

The strongest indication that satellites are subject to unique processes in their outskirts comes from our direct spectral comparison (Figure \ref{fig1}). Without any stellar population characterization, we detected systematic differences in how central and satellite galaxies assemble their stellar components. These differences are magnified beyond the $r_e$ and are caused by variations in [Mg/Fe] (Figure \ref{fig2}), indicating that satellite galaxies assembled their outskirts over remarkably short timescales.

Other patterns observed in MaNGA and SAMI (\citealt{allen2015}) also support an environment-driven, outside-in picture for satellite quenching. The outskirts of star-forming galaxies in high-density regions show evidence of star-formation suppression (\citealt{schaefer2017}) and M$_*$ deficit (\citealt{spindler-wake2017}). In contrast, other works have also reported signatures of \textsl{inside-out} quenching (\citealt{lin2019}), highlighting that further studies of low-mass galaxies are required to better tease out these mechanisms.

\subsection{The stellar accretion scenario}
\label{halo}

Group finding algorithms, including the one used in this paper, match groups in luminosity with dark-matter halos in M$_h$ by their abundance (i.e. abundance matching). This means that if a central and satellite are observed to have the same M$_*$ within the $r_e$ (M$_*^{r_e}$) and their respective groups have the same M$_h$, then their total r-band luminosity (L$_*^r$) must differ. This is highlighted in Figure \ref{fig6}, where centrals and satellites that have been matched in M$_*$ and M$_h$ follow different tracks in M$_*^{r_e}/$L$_*^{r}$.

The lower M$_*^{r_e}/$L$_*^{r}$ of central galaxies can be interpreted in the context of their peculiar surface brightness profiles. While the centers of massive central galaxies ($r<1$ kpc) are very compact, their outskirts feature faint, extended, highly elliptical outer envelopes ($r>10$ kpc) that presumably assembled through the accretion of stellar envelopes from satellite galaxies (\citealt{huang2013a,huang2013b,huang2018}). This picture suggests that differences between the stellar populations of central and satellite galaxies could be due to differences in their accretion histories.

In principle, we could attribute the deficit of [Mg/Fe] in centrals (Figure \ref{fig2}) to the accretion of either Mg-poor or Fe-enriched stellar envelopes. However, given how minor the differences in [Fe/H] between centrals and satellites are (Figure \ref{fig2}), the element driving the differences in [Mg/Fe] must be magnesium. For stellar accretion to explain the differences between the stellar populations of centrals and satellites, we need these accreted stellar envelopes formed to be Mg-poor. 

To test the validity of this picture, we can turn to hydrodynamical simulations. In Illustris (\citealt{rodriguez-gomez2016}), the characteristic M$_*$ difference between a central galaxy and the accreted satellite is $\Delta \log{\mbox{M}_*}\sim 0-0.5$. Since [Mg/Fe] is a monotonically increasing function of M$_*$ (Figure \ref{fig3}), the expectation is that \textsl{ex-situ} stellar populations are more Mg-poor than \textsl{in-situ} stellar populations. This line of thought is consistent with the deficit of [Mg/Fe] in central galaxies from Figure \ref{fig2}.

In the scenario proposed, central galaxies with higher \textsl{ex-situ} mass fractions should have, on average, higher alpha abundances. However, we found the exact opposite in our characterization of central galaxy stellar populations: at fixed M$_*$, central galaxies in more massive dark-matter halos host more alpha enhanced stars (\citealt{oyarzun2022}). Taking into account that centrals in massive halos show stronger merger growth signatures (\citealt{huang2020}), our result casts doubts over whether the stellar population differences between centrals and satellites could be driven by differences in their accretion histories.

It is interesting to consider how systematic variations between the M$_h$ of centrals and the M$_h$ of satellite subhalos (i.e., the M$_h$ prior to infall) could also produce differences in the enrichment of magnesium. Perhaps the halos of central galaxies can efficiently sustain accretion of pristine gas, keeping them relatively Mg-poor. Satellite galaxies, on the other hand, might have assembled as centrals in environments of particularly high density, limiting their ability to sustain cold flows of gas and keeping them Mg-enriched as a result.

\section{Summary}
\label{6}

In this work, we took advantage of the sample size and radial coverage of the MaNGA survey to compute high signal-to-noise stacked spectra for central and satellite galaxies out to 1.5 $r_e$. By using the stellar population fitting code \texttt{alf} and the \citet{tinker2020b} group catalog, we were able to constrain the stellar age, [Fe/H], and [Mg/Fe] profiles of centrals and satellite galaxies as a function of M$_*$, M$_h$, and cluster-centric distance. We found the following:

(1). At fixed M$_*$ and M$_h$, central and satellite galaxies show significant spectral differences at the 1\% level that grow in magnitude and significance as galactocentric distance increases. We associate these differences with variations in [Mg/Fe], with satellite galaxies featuring more alpha enhanced stellar populations. This result reveals that differences between the stellar populations of centrals and satellites of the same M$_*$ cannot be fully ascribed to variations in M$_h$. We consider two scenarios to explain these spectral differences:

\textsl{Environment-driven quenching}: Satellite quenching is facilitated by gravitational and hydrodynamical interactions with their host halos. Star-formation in the outskirts of satellites is inhibited, leading to \textsl{outside-in} quenching signatures in the spatially resolved stellar populations.

\textsl{Stellar accretion}: Central galaxies have built up their outskirts through the accretion of stellar envelopes from satellite galaxies. As a result, the outskirts of centrals contain alpha deficient stellar populations that originally formed in lower M$_*$ satellite galaxies.

(2). The stellar populations of satellite galaxies depend on the mass of the host halo. Satellites in high-M$_h$ halos feature older stellar ages, lower [Fe/H], and higher [Mg/Fe] than satellites in low-M$_h$ halos. A possible explanation for this result is that massive host halos quench satellite galaxies at earlier times (e.g. \citealt{pasquali2010,gallazzi2020}). Alternatively, these results could be driven by how the assembly histories of galaxies vary with the local environment (e.g. \citealt{oyarzun2022}).

\medskip 
We thank everyone at UC Santa Cruz involved in the installation and maintenance of the supercomputer Graymalkin, which was used to run \texttt{alf} on MaNGA spectra. We acknowledge use of the lux supercomputer at UC Santa Cruz, funded by NSF MRI grant AST 1828315. This work made use of GNU Parallel (\citealt{tange2018}). G.O. acknowledges support from the Regents' Fellowship from the University of California, Santa Cruz. K.B. was supported by a UC-MEXUS-CONACYT Grant. R.Y. would like to acknowledge support from the Hong Kong Global STEM Scholar scheme, the Direct Grant of CUHK Faculty of Science, and the Research Grant Council of the Hong Kong Special Administrative Region, China (Project No. 14302522). I.L. acknowledges support from Proyecto DIUDA Programa Inserci\'on No. 22414 of Universidad de Atacama. This research made use of Marvin, a core Python package and web framework for MaNGA data, developed by Brian Cherinka, Jos\'e S\'anchez-Gallego, and Brett Andrews. Funding for SDSS-IV has been provided by the Alfred P. Sloan Foundation, the U.S. Department of Energy Office of Science, and the Participating Institutions. SDSS acknowledges support and resources from the Center for High-Performance Computing at the University of Utah. The SDSS website is \url{https://www.sdss.org}. SDSS is managed by the Astrophysical Research Consortium for the Participating Institutions of the SDSS Collaboration including the Brazilian Participation Group, the Carnegie Institution for Science, Carnegie Mellon University, the Chilean Participation Group, the French Participation Group, Harvard-Smithsonian Center for Astrophysics, Instituto de Astrof\'isica de Canarias, The Johns Hopkins University, Kavli Institute for the Physics and Mathematics of the Universe (IPMU)/University of Tokyo, Lawrence Berkeley National Laboratory, Leibniz Institut f\"ur Astrophysik Potsdam (AIP), Max-Planck-Institut f\"ur Astronomie (MPIA Heidelberg), Max-Planck-Institut f\"ur Astrophysik (MPA Garching), Max-Planck-Institut f\"ur Extraterrestrische Physik (MPE), National Astronomical Observatories of China, New Mexico State University, New York University, University of Notre Dame, Observat\'orio Nacional/MCTI, The Ohio State University, Pennsylvania State University, Shanghai Astronomical Observatory, United Kingdom Participation Group, Universidad Nacional Aut\'onoma de M\'exico, University of Arizona, University of Colorado Boulder, University of Oxford, University of Portsmouth, University of Utah, University of Virginia, University of Washington, University of Wisconsin, Vanderbilt University, and Yale University.

\newpage
\bibliography{sample631}{}
\bibliographystyle{aasjournal}



\end{document}